# On surface plasmon polariton wavepacket dynamics in metal-dielectric heterostructures

**V I Belotelov[1,2], D A Bykov[3], L L Doskolovich[3], A K Zvezdin[1]**

[1]A.M. Prokhorov General Physics Institute of RAS, Vavilov st. 38, Moscow, 119991, Russia
[2]M.V. Lomonosov Moscow State University, Moscow, 119992, Russia
[3]Image Processing Systems Institute RAS, 151, Molodog. st., Samara, 443001, Russia

E-mail: belotelov@physics.msu.ru, zvezdin@gmail.com

**Abstract.** The WKB equations for dynamics of the surface plasmon polariton (SPP) wavepacket are studied. The dispersion law for the SPP in the metal-dielectric heterostructure with varying thickness of a perforated dielectric layer is rigorously calculated and investigated using the scattering matrix method. Two channels of the SPP wavepacket optical losses related to the absorption in a metal and to the SPP leakage are analyzed. It is shown that change of the dielectric layer thickness acts on the SPP as an external force leading to evolution of its quasimomentum and to the wavepacket reversal or even to the optical Bloch oscillations (BO). Properties of these phenomena are investigated and discussed. Typical values of the BO amplitude are about tens of microns and the period is around tens or hundreds of femtoseconds.

## 1. Introduction

Plasmonics is a rapidly developing new field of science and technology studying possibilities of flow of light molding and control at the nanoscale using plasmon polaritons in metal-dielectric nanostructures [1]. Its numerous applications include solar cells with enhanced efficiency, plasmonic nanolithography, negative index metamaterials, optical data storage beyond the diffraction limit, gas and liquid sensing, and plasmonic circuitry [1-2]. The latter is of prime importance for the modern integrated optics devices and optical computing paradigm [3]. Information in the plasmonic circuits is to be conveyed via pulses of SPPs - coupled oscillations of the electromagnetic field and electron plasma propagating along the metal/dielectric interface. The SPPs offer a very promising alternative to carry optical signals in nanoscale networks [4]. That is why investigation of the SPP pulses propagation in different nanostructured systems is very important.

The SPP wavepackets dynamics at smooth metal-dielectric layered structures has been studied in several papers [5-7]. The pulse dispersion and dissipation were studied both numerically and experimentally. The propagation of the SPP pulses in nanostructured media have received much attention with some interesting results [8-14]. Thus, in [9] the authors studied the SPP propagation along a 90° bent line defect in a periodically corrugated metal surface. The propagation of the SPP wavepackets in metal nanoparticle chains immersed in a graded dielectric host was investigated in [10] with focusing on conditions for different types of the SPPs dynamics. Trapping of the SPPs in Bragg structure of a metallic film covered by a dielectric grating of graded thickness is predicted in [14].

One of the interesting optical properties of the SPP wavepackets propagating in periodic nanostructured systems are the optical Bloch oscillations (BOs) and Zener tunneling [15-25]. Actually, BOs were predicted and observed in semiconductors. It describes the oscillation of an electron in the periodic crystal potential if a constant electric field is exerting it [26]. The proportionality of the Bloch frequency to the electric field makes the oscillations tunable, and opens some potential applications for terahertz radiation sources [27] and in Josephson junctions [28]. Though BOs were theoretically predicted in 1928 its experimental observation took place more than half a century later [29] when semiconductor superlattices were fabricated. The reason for that is related to the fact that dephasing time of electrons in conventional crystals is shorter than the period of the BOs and superlatticies are vital to make the oscillations period smaller.

Recently, it was demonstrated that BOs of waves of different nature: electronic, acoustic, optical et al. can appear [15,16,30]. The necessary conditions are the existence of Bragg reflections together with a linear gradient of the potential. Dephasing processes for electromagnetic or acoustic waves are much weaker, and this makes the observation of the Bloch oscillations easier. In the case of optical BOs the optical equivalent of an external field can be realized in specially designed photonic crystal superlattices [17,18], chirped gratings [19] and some other dielectric structures [20,21].

The optical BOs were also studied in metal - dielectric structures where SPPs play a significant role [10,22-25]. These investigations are of prime interest since SPPs allow an alternative to carry optical signals in nanoscale networks and can be used in new SPP-based nanophotonic components and circuits [3]. Most of the research efforts in this area were made for the coupled metal waveguide arrays with one of the geometrical or optical parameters linearly varying in the direction perpendicular to the metal-dielectric interfaces. The SPPs oscillate transversely with respect to their propagation direction [22-25]. It should be noted that an intrinsic drawback of these systems making the experimental observation of the BOs complicated is a considerable dissipation of the electromagnetic energy when SPPs coupled to the adjacent metal-dielectric waveguides. At the same time, authors of [10] investigated plasmonic oscillations in the longitudinal direction by considering the metal nanoparticle chains immersed in a graded dielectric host. Though in this case the BOs take place along the wavepacket propagation, the plasmon polaritons coupling process is still essential but this time it is expressed in the localized plasmon polariton tunneling between the neighboring nanoparticles.

The literary review of the problem reveals that the dynamics of the SPPs pulses in periodic nanostructures systems with varying parameters has been investigated insufficiently leaving many unclear issues. In addition to that a spectrum of the studied systems is far from completeness. A comprehensive investigation of different types of the processes of the SPPs wavepacket dynamics is to be made having in mind its possible applications.

In this paper we are contributing to the above-described research area by investigating the SPPs wavepacket dynamics in a periodically perforated metal-dielectric films having one of the geometrical or optical parameters varying monotonically along the periodicity axis. This situation is significantly different from the aforementioned ones in the sense that it allows continuous propagation of the SPPs wavepacket in the direction of the parameters variation. No tunneling or interfaces coupling is involved here. We start from derivation of the equations for dynamics of the polariton wavepacket that includes the Berry phase effects [31]. These equations are quite similar to the equations for the Bloch electron. Then we proceed with investigation of the SPP wavepacket dynamics in the metal-dielectric gratings. After conducting a general analysis of the problem and formulating necessary conditions for the oscillatory motion we consider one specific example. Namely, we study a system of a smooth metal covered by a dielectric film of monotonically changing thickness perforated with a periodic array of the parallel slits. The parametric dependence of the SPP dispersion law on the dielectric grating thickness is studied rigorously using the scattering matrix approach [32,33]. The calculated dispersion laws for the ideally periodic structure are applied then for solving equations of the wavepacket motion. Different conditions of the SPP wavepacket dynamics are studied including the single reflection, complete oscillatory motion and trapping mode.

The paper is organized as follows. In Sec. 2 we present derivation of the dynamics equations for the motion of the polariton wavepacket center in a periodic system with the presence of effective driving force. Section 3 introduces the general study of the SPPs pulse motion in the quasi periodic systems along with analysis of the parametric dependence of the dispersion law on the dielectric grating thickness. We study SPP oscillatory motion for the case of specific metal-dielectric structure and discuss its main properties and peculiarities in Sec. 4. Section 5 is conclusive.

## 2. Equations of the SPP wavepacket motion

Let us consider an electromagnetic wavepacket propagating along *x*-axis in the structure with a permittivity $\varepsilon(\mathbf{r})$. The permittivity $\varepsilon(\mathbf{r})$ is a quasi-periodic function of *x*-coordinate, i.e. it changes with *x* almost periodically with some small perturbations that are monotonic functions of *x*. The permittivity can be written as a sum of the ideally periodic term $\varepsilon_0(\mathbf{r})$ and a perturbation term $\delta\varepsilon(\mathbf{r})$: $\varepsilon(\mathbf{r}) = \varepsilon_0(\mathbf{r}) + \delta\varepsilon(\mathbf{r})$. The length scale of the perturbation term $\delta\varepsilon(\mathbf{r})$ is much larger than the spatial period of $\varepsilon_0(\mathbf{r})$ and the width of the wavepacket. The structure is stratified in the *z*-direction and is uniform along the y-axis. The permeability of the structure is unity.

It follows from Maxwell equations that magnetic field $\tilde{\mathbf{H}}(\mathbf{r},t)$ of the wavepacket obeys the equation:

$$\nabla \times \frac{1}{\varepsilon(\mathbf{r})} \nabla \times \tilde{\mathbf{H}}(\mathbf{r},t) + \frac{1}{c^2} \frac{\partial^2}{\partial t^2} \tilde{\mathbf{H}}(\mathbf{r},t) = 0, \qquad (1)$$

where *c* is the light velocity in the vacuum.

The magnetic field $\tilde{\mathbf{H}}(\mathbf{r},t)$ of the wavepacket of carrier frequency $\omega_0$ can be presented by

$$\tilde{\mathbf{H}}(\mathbf{r},t) = \mathbf{H}(\mathbf{r},t)\exp(-i\omega_0 t). \qquad (2)$$

It leads to the following equation in the operator form:

$$\left[ i\frac{\partial}{\partial t} - \hat{\mathbf{U}}(\mathbf{r}) \right] \mathbf{H}(\mathbf{r},t) = 0, \qquad (3)$$

where the second time derivative of $\mathbf{H}(\mathbf{r},t)$ is neglected and a Hamiltonian like operator $\hat{\mathbf{U}}(\mathbf{r}) = \frac{c^2}{2\omega_0} \nabla \times \frac{1}{\varepsilon(\mathbf{r})} \nabla \times - \frac{\omega_0}{2}$. This operator is Hermitian providing that $\varepsilon(\mathbf{r})$ is a real quantity or more generally a Hermitian tensor. It should be noted that Eq. (3) is isomorphic to the Schrödinger equation. We assume that the magnetic field amplitude $\mathbf{H}(\mathbf{r},t)$ is normalized, i.e. $\int d^3\mathbf{r} |\mathbf{H}(\mathbf{r},t)|^2 = 1$.

For periodic $\varepsilon(\mathbf{r})$ it is possible to expand the wavepacket onto continuum of the Bloch functions basis:

$$\mathbf{H}(\mathbf{r},t) = \int d^3\kappa\, C(\kappa,t)\, \mathbf{h}_{\kappa,\mathbf{r}_0}(\mathbf{r},t) \qquad (4)$$

with coefficients $C(\kappa,t)$ normalization $\int d^3\kappa |C(\kappa,t)|^2 = 1$. Here $\mathbf{h}_{\kappa,\mathbf{r}_0}(\mathbf{r},t)$ are the Bloch functions of the Bloch wave number $\kappa$ corresponding to the position of the wavepacket's center at the point $\mathbf{r}_0$. Consequently, the wavepacket dynamics is described and can be parameterized in terms of its central position $\mathbf{r}_0(t) = \int d^3\mathbf{r} |\mathbf{H}(\mathbf{r},t)|^2 \mathbf{r}$ and the mean Bloch wave vector $\kappa_0(t) = \int d^3\kappa |C(\kappa,t)|^2 \kappa$.

In our case $\varepsilon(\mathbf{r})$ contains a non-periodic part that can be considered as a perturbation. Since the electromagnetic pulse is spread in space its different parts at any moment are located at different regions of the structure with different quasi-periodic permittivities due to the perturbations $\delta\varepsilon(\mathbf{r})$. However if the length of the wavepacket is much smaller than the length scale of the perturbations then one can linearize the perturbations around the wavepacket's center $\mathbf{r}_0$ and the wavepacket's shape is not important to study its dynamics. Thus the Hamiltonian is approximated by

$$\hat{\mathbf{U}} = \hat{\mathbf{U}}_0 + \Delta\hat{\mathbf{U}}. \qquad (5)$$

The first term considers the pulse propagation in the periodic medium with the permittivity $\varepsilon(\mathbf{r},\mathbf{r}_0) = \varepsilon_0(\mathbf{r}) + \delta\varepsilon(\mathbf{r}_0(t))$ which is ideally periodic but with value slightly changing with time due to the term $\delta\varepsilon(\mathbf{r}_0(t))$. The second term takes into account change of the perturbation $\delta\varepsilon(\mathbf{r})$ and thus it is proportional to $\Delta\hat{\mathbf{U}} \sim (\mathbf{r}-\mathbf{r}_0)\mathbf{grad}\,\delta\varepsilon(\mathbf{r}_0)$.

Since the Hamiltonian $\hat{\mathbf{U}}_0$ is periodic in $\mathbf{r}$, its eigenvalue problem is given by:

$$\hat{\mathbf{U}}_0(\mathbf{r},\mathbf{r}_0)\mathbf{h}_{\kappa,\mathbf{r}_0}(\mathbf{r},t) = \delta\omega(\kappa,\mathbf{r}_0)\mathbf{h}_{\kappa,\mathbf{r}_0}(\mathbf{r},t), \qquad (6)$$

where $\delta\omega(\kappa,\mathbf{r}_0) = \omega(\kappa,\mathbf{r}_0) - \omega_0$ is the Bloch function frequency. We work in the single band approximation so the band index is omitted.

Further derivation can be performed in the framework of the time dependent variational principle $\delta S = 0$, where $S = \int L dt$ is the action of the SPP field. The Lagrangian of the system corresponding to (3) is given by

$$L = \int d^3\mathbf{r}\left[i\mathbf{H}^*(\mathbf{r},t)\frac{d}{dt}\mathbf{H}(\mathbf{r},t) - \mathbf{H}^*(\mathbf{r},t)\hat{U}(\mathbf{r})\mathbf{H}(\mathbf{r},t)\right]. \tag{7}$$

The Lagrangian is found as a function of $\mathbf{r}_0$, $\dot{\mathbf{r}}_0$, $\mathbf{\kappa}_0$, and $\dot{\mathbf{\kappa}}_0$. Substituting the field $\mathbf{H}(\mathbf{r},t)$ in the form of (4) in (7) and taking into account (5) and (6) one can present the Lagrangian as (in a close analogy with the Lagrangian of the electron wavepacket in a crystal [34])

$$L = -\delta\omega + \mathbf{\kappa}_0\dot{\mathbf{r}}_0 + \langle u | i\, du/dt \rangle, \tag{8}$$

where $\mathbf{u} = \mathbf{u}(\mathbf{r}_0,\mathbf{\kappa}_0,t)$ is the Bloch periodic amplitude. It should be noted that the last term of the Lagrangian represents the well-know formula for the Berry phase [31] for the wavepacket motion within the band. Using Lagrangian (8) one gets the following Euler–Lagrange equations of the wavepacket motion:

$$\dot{\mathbf{\kappa}}_0 = -\frac{\partial\omega}{\partial\mathbf{r}_0} + \left(\hat{T}_{\mathbf{rr}}\dot{\mathbf{r}}_0 + \hat{T}_{\mathbf{r\kappa}}\dot{\mathbf{\kappa}}_0\right) - \hat{T}_{t\mathbf{r}}$$

$$\dot{\mathbf{r}}_0 = \frac{\partial\omega}{\partial\mathbf{\kappa}_0} - \left(\hat{T}_{\mathbf{\kappa r}}\dot{\mathbf{r}}_0 + \hat{T}_{\mathbf{\kappa\kappa}}\dot{\mathbf{\kappa}}_0\right) + \hat{T}_{t\mathbf{\kappa}}$$
, (9)

where tensors $\left(\hat{T}_{\mathbf{fg}}\right)_{\alpha\beta} = i\left[\left\langle\frac{\partial u}{\partial f_\alpha}\bigg|\frac{\partial u}{\partial g_\beta}\right\rangle - \left\langle\frac{\partial u}{\partial g_\beta}\bigg|\frac{\partial u}{\partial f_\alpha}\right\rangle\right]$, $\left(\hat{T}_{t\mathbf{f}}\right)_\alpha = i\left[\left\langle\frac{\partial u}{\partial t}\bigg|\frac{\partial u}{\partial f_\alpha}\right\rangle - \left\langle\frac{\partial u}{\partial f_\alpha}\bigg|\frac{\partial u}{\partial t}\right\rangle\right]$,

$\mathbf{f}, \mathbf{g} = \mathbf{r}, \mathbf{\kappa}$, and $\alpha, \beta = x, y, z$. Tensors $\hat{T}_{\mathbf{fg}}$ and $\hat{T}_{t\mathbf{f}}$ are called Berry curvatures [31].

The occurrence of the Berry curvatures takes place for special kinds of SPP dispersion bands, e.g. $\omega(\mathbf{\kappa}) \neq \omega(-\mathbf{\kappa})$. Such a behavior can be present in magnetic fields [31] or for a special chiral plasmonic film. In this work we consider the case when Berry curvatures equal to zero and Eqs. (9) are simplified to have only the first terms in their right hands. Actually, Eqs.(9) thus simplified are generalization for SPP of the equation for the dynamics of the electron Bloch wave vector $\mathbf{k}$ in an external electric field $\mathbf{E}$: $\hbar\dot{\mathbf{k}} = -e\mathbf{E}$ in which $\frac{\partial\omega}{\partial\mathbf{r}_0}$ plays the role of the external force acting on the photon wave packet. As we already mentioned into introduction part, the presence of the external force can lead to the Bloch oscillations of the electron. Consequently, one can expect that similar oscillations of the electromagnetic wavepacket appear.

**3. Movement and Bloch oscillations of the wavepacket with a spatially dependent dispersion law**
It follows from (9) that there are at least two necessary conditions for the oscillatory motion of the wavepacket. These are the presence of the structure periodicity and the existence of the "external force". The latter can be generated by making one of the geometrical or optical parameters of the structure be dependent on the spatial coordinates. In this work we investigate dynamics of the plasmonic wavepacket. Consequently, metal-dielectric structures supporting SPPs should be considered. A metal-dielectric perforated film with varying geometrical or optical parameters is a quite acceptable candidate for observation of the SPP wavepacket oscillatory motion. In this case variation of the grating period, film thickness, permittivity of the dielectric or plasma frequency of the metal can be used to set up the external force. Let us consider a structure consisting of a thick smooth metal film covered with a dielectric layer periodically perforated (the period is *d*) with an array of parallel slits of width *r*, the thickness of the dielectric layer changes linearly along the direction perpendicular to the slits (Fig.1). Due to the spatial variation of the dielectric grating thickness the dispersion of the SPPs also changes with *x*-coordinate. Thus, the SPPs pulse when propagating along the interface between the metal film and the dielectric grating feels variation in the local dispersion law and consequently it experiences the "force".

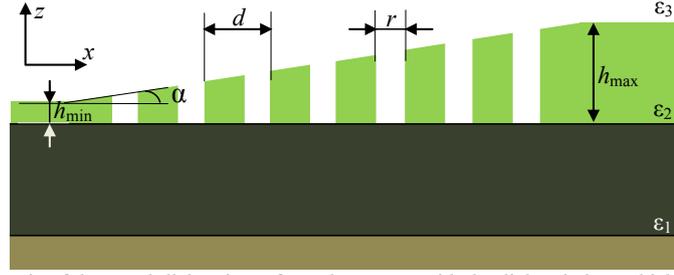

Fig. 1 Schematic of the metal-dielectric perforated structure with the dielectric layer thickness linearly varying with *x*-coordinate.

Let us start from the case of the metal-dielectric system with the dielectric wedge without any perforation. Dispersion of the SPPs in the smooth metal-dielectric film of constant height is found by solving the following transcendental equation [35]:

$$\eta_2(\eta_1 + \eta_3) + [\eta_1\eta_3 + \eta_2^2]\tanh(\gamma_2 h) = 0, \qquad (10)$$

where $\eta_i = \gamma_i/\varepsilon_i$, $\gamma_i = \sqrt{\kappa^2 - \varepsilon_i k_0^2}$, $k_0 = \omega/c$, $h$ is the thickness of the dielectric film, $\varepsilon_i$ are the permittivities of the metal ($i=1$), of the dielectric film ($i=2$), and of the dielectric superstrate which is air in our case ($i=3$). While the permittivities of both dielectric media can be considered constant, the metal permittivity $\varepsilon_1 = \varepsilon_1' + i\varepsilon_1''$ is essentially frequency dependent and is well described by the Drude-Sommerfeld model. There are two characteristic frequencies of the system: $\widetilde{\omega}_{12}$ and $\widetilde{\omega}_{13}$ given by: $\varepsilon_1'(\widetilde{\omega}_{12}) + \varepsilon_2 = 0$ and $\varepsilon_1'(\widetilde{\omega}_{13}) + \varepsilon_3 = 0$. These frequencies are the maximum possible frequencies of the SPPs excitation at the interface between the metal and with either the first or the second dielectric. If $\varepsilon_2 > \varepsilon_3$, then $\widetilde{\omega}_{12} < \widetilde{\omega}_{13}$. It can be shown that for frequencies $\omega < \widetilde{\omega}_{12}$ the dependence of the SPPs dispersion on the thickness $h$ is monotonic implying that for relatively small values of $h$ the dispersion curve $\omega(\kappa)$ is close to the dispersion curve of the SPP at the interface between the metal and air $\omega_{13}(\kappa)$ and for sufficiently large values of $h$ the dispersion curve $\omega(\kappa)$ is close to the dispersion curve of the SPP at the interface between the metal and the dielectric $\omega_{12}(\kappa)$ while for the intermediate values of $h$ it is somewhere in between: $\omega_{12}(\kappa) < \omega(\kappa) < \omega_{13}(\kappa)$ [35]. The dispersion curves that follow from (10) for the metal/dielectric/air trilayer corresponding to different thicknesses of the dielectric layer are shown in Fig. 2.

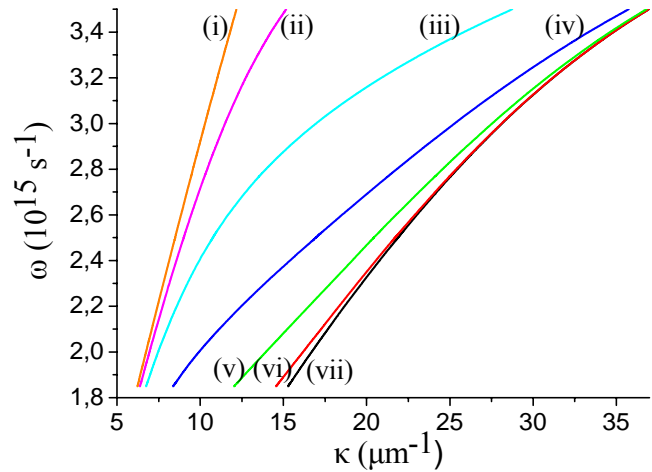

Fig. 2 Dispersion curves of the SPP in the silver/dielectric/air system at different thicknesses *h* of the dielectric layer ($\varepsilon_2 = 5.5$): (i) 0; (ii) 20 nm; (iii) 40 nm; (iv) 70 nm; (v) 110 nm; (vi) 200 nm; (vii) 600 nm.

At the chosen frequency interval the dispersion is sensitive to the dielectric thickness for the thicknesses less than 230 nm. Since variation of the dielectric layer thickness modifies the SPP dispersion law, the "force" will act on the SPP wavepacket propagating in a metal-dielectric system with a smooth dielectric wedge as long as $h$ is less than 230 nm. So, the SPP wavepacket acceleration in a non-perforated system is present. However, an oscillatory motion is not possible since there is not any periodicity.

The presence of the periodicity leads to some modifications of the dispersion law for the smooth system. The wave vector becomes a quasi-wave-vector or quasimomentum and the dispersion is conveniently described in the first Brillouin zone. In addition to that splittings at the edges of the Brillouin zone appear. While rigorous calculation of the SPP dispersion law in the perforated case we defer until the next chapter we study here an artificially simplified situation when the dispersion law $\omega(\kappa, x)$ is expressed by retaining only the first two terms in the Taylor series:

$$\omega(\kappa, x(t)) = \omega(\kappa, x(0)) + b(x(t) - x(0)), \tag{11}$$

where $x(t)$ and $x(0)$ are current and initial coordinates of the wavepacket center, and $b = \frac{\partial \omega}{\partial x}(x(0))$. Though this approximation is usually valid only for a very narrow range of $x$, it is instructive to start with it since it allows getting an analytical solution of (9):

$$\kappa_0(t) = \kappa_0(0) - bt;$$
$$x_0(t) = \frac{1}{b}[\omega(\kappa_0(0), x_0(0)) - \omega(\kappa_0(t), x_0(0))], \tag{12}$$

Arguing in terms of the first Brillouin zone the quasi-wave-vector $\kappa$ - varies periodically in the interval $[-\pi/d; \pi/d]$, where $d$ is the period of the structure: when it approaches $\pi/d$ it jumps back to $-\pi/d$ and continues its increase. It is worth recalling that the use of Eq.(9) implies working in a single band approximation when the wavepacket transfer into neighboring bands is negligibly small. Having this in mind it is straightforward to find that the wavepacket center $x(t)$ oscillates with an amplitude of

$$\Delta x = \frac{1}{2b}[\omega(0, x_0(0)) - \omega(\pi/d, x_0(0))], \tag{13}$$

and a period of

$$T = \frac{2\pi}{bd}. \tag{14}$$

For the SPP propagation in a metal-dielectric heterostructure some typical values of the BO amplitude are about several microns or tens of microns and the period is around tens or hundreds of femtoseconds. At this, the group velocity of the wavepacket motion between two turning points can be estimated as $10^8 \, m/s$. The turning points at which the SPP pulse changes direction of its motion to the opposite one correspond to the quasimomentum values of $\kappa = 0$ and $\kappa = \pi/d$. The necessary condition for the wavepackets spatial reversal is that the wavepacket's group velocity $\frac{\partial \omega}{\partial \kappa}$ is zero. At this, the central frequency of the wavepacket remains constant, i.e. its trajectory in the $\omega(\kappa)$ diagram is a horizontal straight line. In the most of practical cases the situation is more complex since expansion (11) is not always valid for the distances compared to the BOs amplitude and $b$ does depend on $\kappa$. Nevertheless, aforementioned expressions for the BOs parameters are still proper for estimations.

**4. Surface plasmon at the perforated metal-dielectric film**
Dispersion law of the SPP at the given $x$-coordinate of the perforated structure shown in Fig. 1 is determined by local height of the dielectric grating. That is why, to investigate propagation of the SPPs wavepacket in this system one needs to calculate SPPs dispersion dependence on the height of the perforated dielectric film. Since the pulse propagates in the periodic structure its dispersion law is reduced to the domain of $\kappa$ lying in the first Brillouin zone. At the edges of the first Brillouin zone the band gaps appear, the width of the gaps being dependant on the optical contrast of the grating and on the width of its slits. Some additional band gaps are also possible at Γ point of the Brillouin zone corresponding to zero $\kappa$.

There are several approaches to the calculation of the SPP dispersion in periodic structures. Some of them are approximate and allow qualitative analysis only [14]. One of the most efficient ways to calculate eigenmodes dispersion is the scattering matrix (S-matrix) method [31,32,34]. The S-matrix approach allows calculation of the eigenfrequencies of the systems with a rather complex structure. While the details of the S-matrix method can be found elsewhere [31,32,34] we mention here that the S-matrix relates complex amplitude of the incident waves and waves scattered by the structure. Eigenmodes of the structure can propagate in it without any external excitation, i.e. they are solutions of the homogeneous problem with zero incident waves. Consequently, at the eigenfrequencies of the structure a determinant of the S-matrix inverse must be zero. Thus, the essence of the S-matrix method is to construct the S-matrix of the system numerically and to find zeros of the determinant of its inverse:

$$\det(\mathbf{S}^{-1}) = 0 \qquad (15)$$

The S-matrix is a function of frequency and quasimomentum $\kappa$, so Eq. (15) can be solved in two different ways: one can assume that the frequency is determined and find complex valued solutions for quasimomentum $\tilde{\kappa}$ or otherwise one can consider the quasimomentum to be real and known and solve (15) for complex valued frequency. The choice between these two alternatives depends on the problem to be resolved. If the propagation of the wavepacket is considered then processes of its dissipation along the trajectory are inevitable and the assumption that it is quasimomentum $\tilde{\kappa}$ which is complex and to be found is more adequate. Since in the present work we study the SPP pulse propagation we chose the first approach and present the quasimomentum as a sum of its real and imaginary parts: $\tilde{\kappa} = \kappa + i\kappa''$.

The dispersion diagram for the metal-dielectric structure of period $d$=280 nm and with slits of the width $r$=60 nm calculated using the S-matrix method is shown in Fig.3. Dispersion curves corresponding to the SPP modes at 5 gratings with different dielectric thicknesses $h$ ranging from 5 nm to 230 nm are presented in Fig. 3a. The strongest dispersion dependence on $h$ takes place for $h$<70 nm and for higher $h$ it weakens rapidly to almost vanish for $h$>230 nm. Consequently, in this case the operating range of $h$ spreads from about 5 nm to 230 nm. From consideration of the simplified case of constant derivative $\frac{\partial \omega(\kappa, x)}{\partial x}$ we found out that frequency of the SPP wavepacket remains steady during its motion trough the graded structure, i.e. a trajectory of the wavepacket at the dispersion diagram is a straight line parallel to κ-axis (thin horizontal lines in Fig. 3). Since in the considered metal-dielectric heterostructure the derivative $\frac{\partial \omega}{\partial h}$ is negative the quasimomentum $\kappa$ increases with time and the wavepacket moves in $(\omega, \kappa)$ space to the positive direction of $\kappa$-axis. The necessary condition for the oscillatory motion is that the SPP pulse position in the dispersion diagram is always inside the region of the driving force action.

The first plasmonic bands for each thickness are shown with dashed lines. The region between two extreme lines corresponding to $h$ =5 nm (black line) and $h$ =230 nm (magenta line) does not cover all range of the quasimomentum in the Brillouin zone. Consequently, complete oscillations of the wavepacket are not possible. Nevertheless, one wavepacket reverse can happen as long as the SPP frequency is in the interval between $\omega_a$ and $\omega_b$ (See Fig 3.a).

The second plasmonic bands are shown with solid lines. They differ from the first bands significantly in the sense that there is a frequency range between $\omega_c$ and $\omega_b$ (see Fig.3b) in which the whole interval of $\kappa$ in the Brillouin zone lies between two extreme lines. For these frequencies complete BOs are allowed.

The other issue to be observed in Fig.3 is the SPPs band gaps taking place at the edges of the Brillouin zone. Their values also depend on the thickness $h$. The band gap width is a very important parameter for the wavepacket dynamics since if it is sufficiently small a SPPs tunneling into neighboring bands is possible – the phenomenon similar to Zener tunneling in the solid state physics and in optics [37,38]. As we discussed above there is one more point at $\kappa$ space crucial for the cycling motion of the pulse – it is the point at $\kappa = 0$. The wavepacket turns around at this point. However, for the metal-dielectric structure considered here the corresponding band gap is absent which makes working in a single band approximation not quite correct, since the wavepacket transfer into neighboring bands is inevitable.

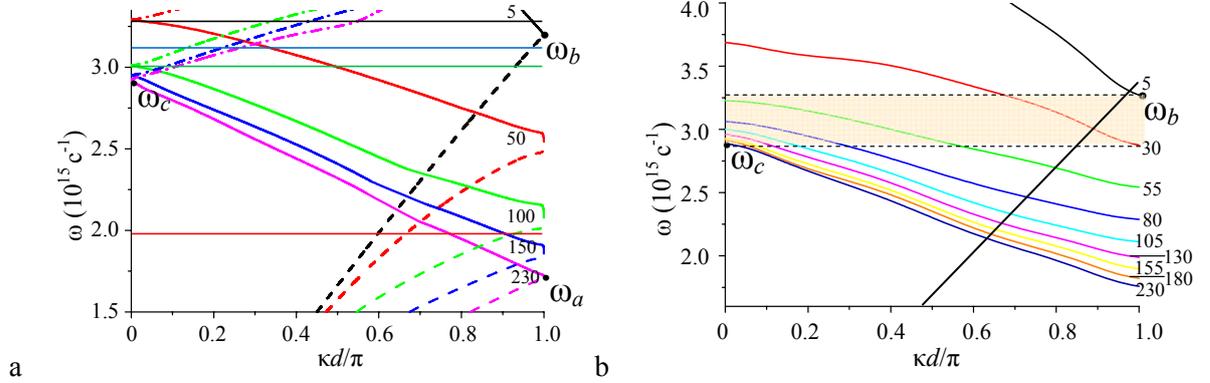

Fig. 3 Dispersion curves $\omega(\kappa)$ of the metal-dielectric heterostructure with $r$=60 nm, $d$=280 nm, and different thicknesses $h$ as indicated on the right side. (a) three plasmonic band are shown with dashed lines (1-st band), solid lines (2$^{nd}$ band) and dash-dotted lines (3$^{rd}$ band). Horizontal colored lines are SPP trajectories at its movement through the structures of varying thickness. (b) More detailed view of the 2$^{nd}$ band. Two horizontal dashed lines are boarders of the region in which BOs are allowed. Black solid line is a boarder of the light cone.

The force is actually characterized by $\frac{\partial \omega}{\partial x}(x,\kappa) = \frac{\partial \omega}{\partial h}(x,\kappa)\tan\alpha$, where α is the angle of the dielectric wedge ascent. The angle α should be chosen so that the time period of the BOs is much larger than the period of high frequency oscillations of the SPP wavepacket carrier: $T \gg 2\pi/\omega$. Estimation of the BOs period by the one for the "constant force" case expressed by (14) gives $\tan\alpha \ll \omega\left(\frac{\partial \omega}{\partial h}d\right)^{-1}$. On the other hand the angle α cannot be too small since otherwise the spatial period of the BOs $\Delta x$ gets too large. The limitation of the spatial period arises from inevitable losses of the SPPs wavepacket energy happening during its propagation along the structure. The upper limit for the angle α is estimated using (13) as $\tan\alpha \gg \Delta\omega\left(\frac{\partial \omega}{\partial h}L\right)^{-1}$, where $\Delta\omega$ is the difference between maximum and minimum values of $\omega$ in the given dispersion band $\omega(\kappa)$ and $L$ is the propagation length of the SPP, i.e. it is a distance at which SPPs energy decreases by a factor of $e$. The propagation length $L$ is given by $L = \frac{1}{2\kappa''}$. The value of $\kappa''$ for two structures with perforated dielectric layers of thicknesses $h$=50 nm and 100 nm are presented in Fig. 4. In both cases curves corresponding to the first band (dashed lines) have rather small absorption, while the curves belonging to the second band contain two parts with relatively small and large $\kappa''$. The curves from the third band have large $\kappa''$ only. To understand such situation we need to take into account that there are two main channels of the SPPs energy decay. The first one is an oscillations damping in a metal and in an absorptive dielectric. The maximum propagation length $L$ can be estimated from considering the SPP on a smooth interface by

$$L = \frac{\lambda}{2\pi}\frac{(\varepsilon_1')^2}{\varepsilon_1''}\left(\frac{\varepsilon_1' + \varepsilon_2}{\varepsilon_1'\varepsilon_2}\right)^{3/2}. \qquad (16)$$

For the silver/dielectric($\varepsilon_2$=5.5) interface Eq.(16) estimates $L$ to be about 40 μm at $\omega = 1.96\cdot10^{15}$ s$^{-1}$. The S-matrix gives $\kappa'' = 0.0007 k_0$ at $\omega = 1.96\cdot10^{15}$ s$^{-1}$ corresponding to the first Brillouin zone and $h$=250 nm. It means that the propagation length $L = 110$ μm which is consistent with the one for the smooth silver/dielectric interface if we take into account finite thickness of the dielectric layer and the presence of the slits.

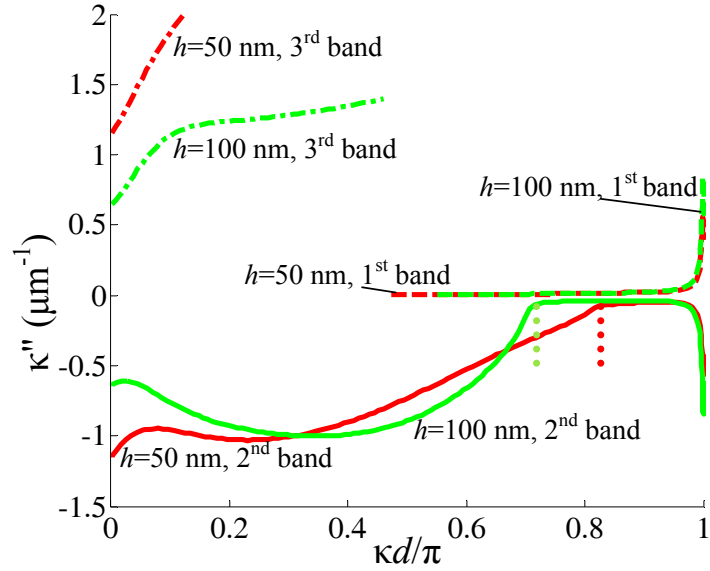

Fig. 4 Imaginary part of the quasimomentum $\kappa''$ versus its real part $\kappa$ for the structures with perforated dielectric layer of thicknesses $h$=50 nm (red curves) and $h$=100 nm (green curves). Three plasmonic bands are shown with dashed, solid and dash-dotted lines. Vertical dotted lines indicate $\kappa$ at which the dispersive curves intersect with the edge of the light cone.

The second way for the SPP pulse decay is the SPPs leakage into the propagating diffraction orders due to the scattering on the dielectric periodic structure. Such SPP out-coupling back into free photons becomes possible at those parts of the dispersion curve which lie inside the light cone. The second SPP bands are partly inside the light cone and partly outside of it. If the SPP wavepacket leakage takes place then $\kappa''$ gets substantially large. It is observed clearly for the second plasmonic bands. For the dielectric grating height $h$=50 nm at $\omega = 2.91 \cdot 10^{15}$ s$^{-1}$ the SPP radiation happens leading to $\kappa'' = 0.12 k_0$ and $L = 450$ nm. However for the same dielectric grating thickness at $\omega = 2.65 \cdot 10^{15}$ s$^{-1}$ the SPP belonging to the second band does not radiate and its decay is only due to the optical absorption of silver with $\kappa'' = 0.0046 k_0$.

The inclination angle α in this case can be chosen to be equal to several degrees so that the duration of the BOs is substantially larger than the period of one oscillation and smaller than the time at which wavepacket's energy dissipates.

Once general conditions and properties of BOs are discussed let us now investigate the process in details. For this we solve Eq. (9) working in the single band approximation and retaining only first terms in the right hand of both equations. The former is well justified for the first band, while for the second band it is not absolutely correct since there is no splitting at Γ point ($\kappa = 0$). Nevertheless, a semiquantitative analysis is still possible even for this case if bearing in mind that at Γ point the wavepacket divides into two parts: one travelling in the same band further and the other one transferring to the neighboring band. Fig.5 presents the SPPs wavepacket motion for several cases of different initial conditions $x_0(0)$ and $\kappa_0(0)$ and the wavepacket central frequencies in the structure with parameters given above. Curve-(a) corresponds to the SPPs wavepacket from the first plasmonic band. Its trajectory in $(\kappa, \omega)$ space is shown in Fig.3 with a thin horizontal red line. The pulse starts at $h(0) = 10$ nm and travels decelerating to the thicker part of the dielectric grating until its quasimomentum reaches the edge of the first Brillouin zone ($\kappa = \pi/d$) where the group velocity equals zero. At this moment the wavepackets turns around and continues motion in opposite direction. In the representation of the reduced Brillouin zone the quasimomentum changes by $-2\pi/d$ at this point and it continues its ascending till the next reduction. The wavepacket after its reversal accelerates up to the point with $h = 5$ nm where the thickness of the grating stops changing with coordinate and consequently the driving force vanishes and the wavepacket travels further at a constant group velocity. The dissipation of the wavepacket energy during this motion is purely determined by the optical absorption in the metal. Since the SPP belongs to the first plasmonic band no energy radiation into far field happens. That is why its propagation distance is about 110 μm which allows it to make the whole round and even go further till it dissipates substantially. Since minimum frequency of any dispersion curves from the first band is zero complete BOs at this band are not possible. The wavepacket can make only one turn around at most.

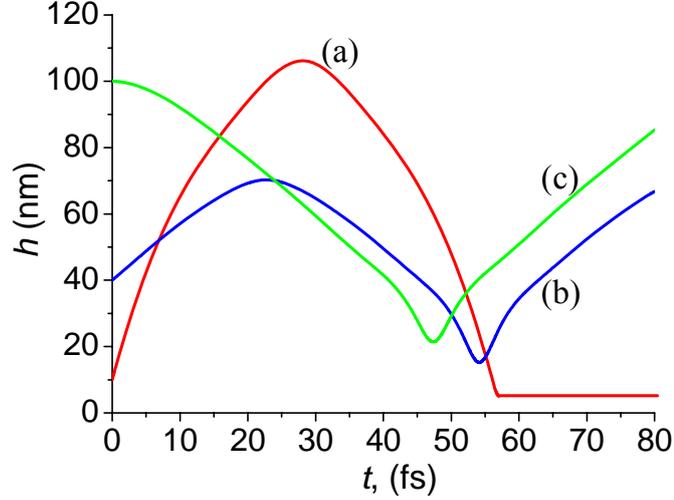

Fig. 5 The SPP wavepacket dynamics: position of the wavepacket center (expressed via local thickness of the dielectric grating, $x_0(t) = (h(t) - h_{min})\cot\alpha$, where $h_{min} = 5$ nm) versus time. Three different cases of initial conditions are considered: (a) $\kappa(0) = 0.6\pi/d$, $h(0) = 10$ nm, $\omega = 1.98 \cdot 10^{15}$ s$^{-1}$; (b) $\kappa(0) = -0.6\pi/d$, $h(0) = 40$ nm, $\omega = 3.11 \cdot 10^{15}$ s$^{-1}$; (c) $\kappa(0) = 0$, $h(0) = 100$ nm, $\omega = 3.0 \cdot 10^{15}$ s$^{-1}$. Parameters of the metal-dielectric heterostructure are the same as in Fig. 3. Inclination angle $\alpha = 1.5°$.

Let us now consider several cases of the propagation of the SPPs pulse belonging to the second plasmonic band. Curve-(b) from Fig. 5 represents the situation when $\kappa(0) < 0$. The wavepacket initially travels to thicker parts of the dielectric and after $\kappa(t)$ becomes zero it turns back and travels to the thinner parts of the grating till its quasimomentum reaches the edge of the first Brillouin zone. At $\kappa(t) = \pi/d$ the wavepacket turns around again and proceeds to its initial position. The BOs phenomenon is fully present. The wavepacket makes about one complete round before its energy decreases significantly. The amplitude of the wavepacket gets sin smaller The corresponding wavepacket trajectory in the $\omega(\kappa)$ diagram is shown in Fig.3 with thin horizontal blue line.

If $\kappa(0) \geq 0$ then the wavepacket starts traveling to the thinner parts of the dielectric grating from the very beginning and changes its motion direction to the opposite one at the edge of the Brillouin zone (see curve-(c) in Fig.5 and green horizontal line in Fig.3). The thicknesses of the dielectric grating at which the wavepacket makes its turns depend on the initial conditions also. As we have already discussed the dissipation of the wavepacket energy in the second band is rather high because of the SPP continuous radiation in the far field at the most of the traveling distance when it is inside the light cone. The propagation length is estimated to be around 1 micron. So the angle of the dielectric wedge should be sufficiently large (about 8°) to allow SPP wavepacket make one whole round before its full dissipation.

If at $\kappa(0) = 0$ $h(0) < 52$ nm then the BOs do not happen since the wavepacket reaches critical point with $h=5$ nm earlier than it its quasimomentum κ reaches $\pi/d$. In this case the trajectory is like the one shown in Fig. 3 by a horizontal black line.

It is noticeable that the times of the wavepacket's turns at two opposite turning points are quite different. Thus, the wavepacket spends much longer at the thicker regions of the dielectric grating. It can be explained by smaller values of the driving force $\frac{\partial\omega}{\partial h}$ being compared to the one for thinner parts of the gratings. In principal, if the wavepacket starts from the parts of the grating a little bit thicker than 170 nm where the $\frac{\partial\omega}{\partial h}$ is not exactly zero but very close to it, then the wavepacket will spend larger fraction of time in this region of grating. If the losses related to the SPPs scattering are high enough then the wavepacket can even loose all its energy there and no further turn occurs. This phenomenon was described recently in [14] where the authors report the possibility of the light of the given frequency trapping in a small region of the grating of varying thickness. Since in real situation the optical losses of the SPP wavepacket are inevitable one more necessary condition for the BOs to take place is that the driving force should stay sufficiently large for the whole pass of the pulse. Otherwise it fully dissipates even before its first turn.

## 5. Conclusion

We studied propagation of the SPP pulse in the perforated metal-dielectric heterostructure with graded optical properties by solving the WKB equations of motion of the wavepacket center. The dispersion law of the SPP in the structure is calculated by finding poles of the S-matrix of the structure. The case of the constantly changing thickness of the dielectric grating is studied in details. We investigated the wavepacket dynamics paying most attention to the possibilities of its reverse and BOs. Conditions necessary for such behavior of the wavepacket are revealed.

The BOs described here can be manifested both in the near and far fields. For the former they are of prime importance for the localization of light of given frequency since the presence of the SPPs wavepackets oscillating in a small region of the nanostructured system leads to several important phenomena related for example to a significant electromagnetic energy concentration and increase of some nonlinear or magnetooptical effects. The appearance of the BOs should also be observed in the far field if the wavepacket belongs to the second or higher plasmonic bands because of the SPPs leakage during its oscillations. One can expect for example a periodic change in position and diffraction efficiency of the diffraction orders in reflected light which implies many new fascinating applications of the optical BOs for light control and optical data processing. Thus the SPPs leakage plays a twofold role. On the one hand it makes the propagation distance much shorter and only several oscillations if any happen. On the other hand SPPs radiation is important for the SPP oscillations observation in the far field. Consequently, a compromise is to be found. The level of the SPP leakage can be controlled via change of the stripe size in the dielectric grating. Meanwhile, some other plasmonic structures with diminished leakage and thus increased SPP propagation distances are to be found and investigated.

**Acknowledgements**
Work is supported by RFBR (08-02-00717, 09-02-01028, 09-02-92671, 10-02-01391, 10-02-91170).